# Machine Learning-Assisted Least Loaded Routing to Improve Performance of Circuit-Switched Networks

Gangxiang Shen, *Senior Member, IEEE*, Longfei Li, Ya Zhang, Wei Chen, Sanjay K. Bose, *Senior Member, IEEE*, Moshe Zukerman, *Fellow, IEEE*

*Abstract*—The Least Loaded (LL) routing algorithm has been in recent decades the routing method of choice in circuit switched networks and therefore it provides a benchmark against which new methods can be compared. This paper improves the performance of the LL algorithm by additionally incorporating a machine learning approach, using a conceptually simple supervised naïve Bayes (NB) classifier. Based on a sequence of historical network snapshots, this predicts the potential future circuit blocking probability between each node pair. These snapshots are taken for each service request arriving to the network and record the number of busy capacity units on each link at that instant. The candidate route for serving a current service request is based on both the link loads and the potential future blocking probability of the entire network in case this route is indeed used. The performance of this proposed approach is studied via simulations and compared with both the conventional LL algorithm and the Shortest Path (SP) based approach. Results indicate that the proposed supervised naïve Bayes classifier-assisted LL routing algorithm significantly reduces blocking probability of service connection requests and outperforms both the conventional LL and SP routing algorithms. To enable the learning process based on a large number of network snapshots, we also develop a parallel computing framework to implement parallel learning and performance evaluation. Also, a network control system supporting naïve Bayes classifier-assisted LL routing algorithm is addressed.

## I. INTRODUCTION

ROUTING algorithms are essential for proper and efficient operation of circuit-switched networks and have been studied for many years. In general, these can be categorized as fixed shortest path routing [1-6], fixed-alternate path routing

This work was jointly supported by National Natural Science Foundation of China (NSFC) (61671313) and the Science and Technology Achievement Transformation Project of Jiangsu Province, PR China (BA2016123).
G. Shen, L. Li, and Y. Zhang are with the School of Electronic and Information Engineering, Soochow University, Suzhou, Jiangsu Province, P.R. China, 215006 (phone: 86-512-65221537; fax: 86-512-65221537; corresponding e-mail: Gangxiang Shen <shengx@suda.edu.cn>). W. Chen is with Key Lab. of New Fiber Tech. of Suzhou City, Jiangsu Hengtong Fiber Science and Technology Corporation, China. S. K. Bose is with the Dept. of EEE, IIT Guwahati, India. M. Zukerman is with the Dept. of EEE, City University of Hong Kong.

[7-10], and adaptive routing [11-14]. Fixed shortest path routing always chooses the fixed shortest route between a pair of nodes to establish the service connection. Fixed-alternate path routing works with a set of routes for service establishment, rather than just one. These are tried in sequence for route establishment until all the routes have been tried. Adaptive routing, also referred to as online routing, does not have a set of predetermined routes. Instead, it chooses the most efficient route for each service demand based on the current network status, e.g., resource utilization on each link. Of these, the Least Loaded (LL) routing algorithm (or least congested routing algorithm) [15-19] has so far been the most efficient algorithm exhibiting the lowest overall average connection blocking probability and has remained so for the past three decades. This paper tries to improve on this for the first time by incorporating a machine learning based approach to LL route selection, which additionally tries to reduce potential future blocking as well.

### A. Machine Learning Techniques in Communication Networks

Machine learning (ML) techniques have become popular in many applications because they can provide frameworks for solving difficult problems. They have also been used to solve optimization problems in telecommunication networks. For example, for optical networks, Huang *et al.* [20] presented a set of intelligent pre-adjustment strategies enabled by machine learning to tackle spectrum defragmentation. Ohba *et al.* [21] applied Bayesian inference to a virtual network reconfiguration framework and introduce the Bayesian Attractor Model to infer the current traffic situation. Barletta *et al.* [22] used a random forest to predict the probability of a candidate lightpath. Morales *et al.* [23] proposed a flow controller to allow metro controllers to share metro-flow predictive traffic models with the core controller. Chen *et al.* [24] proposed a knowledge-based autonomous service provisioning framework enabled by a deep neural network-based traffic estimator for broker-based multi-domain software-defined elastic optical networks. Samadi *et al.* [25] proposed a cognitive scalable method based on neural networks to address dynamic and agile provisioning of optical physical layer without prior knowledge of network specifications. Meng *et al.* [26] demonstrated a self-learning network with dynamic abstraction process based on real-time monitoring and Markov chain Monte Carlo simulations. Similarly, in wireless networks, Forster [27] presented a survey on the usage of machine learning techniques for data routing in wireless sensor networks (WSNs) and mobile ad-hoc networks (MANETs). Russell *et al.* [28] presented an improved wireless adaptive routing protocol using machine learning techniques. Lee *et al.* [29] applied reinforcement learning for wireless network management to reduce protocol overhead and improve the packet delivery ratio.

ML techniques have also long been used in the circuit-switched networks. Boyan and Littman [30] first applied the reinforcement learning technique for routing in circuit-switched networks. They presented a Q-routing algorithm, which discovers the efficient routing policies in a dynamically changing network. Choi and Yeung [31] subsequently extended to propose a gradient algorithm based on Q-routing. Recently, Li *et al.* [32] also employed a technique similar to reinforcement learning to train a set of fixed routes that are the best to serve service connections in circuit-switched optical networks. Leung *et al.* [33] proposed the neural network approach for the blocking probability evaluation on optical networks, which can greatly accelerate its evaluation speed of the blocking probability. However, these circuit-switched routing algorithms are found to be fixed route oriented, which would significantly limit overall routing performance due to their inflexibility in route selection when provisioning online service connections.

*B. Summary and Our Contributions*

In circuit-switched networks, for the past several decades, the LL routing algorithm has been used as a benchmark because it generally does best in terms of lower connection blocking probability. Even though ML techniques have been tried in various areas, including communication networks, to the best of our knowledge, currently no ML technique has been applied to enhance routing performance for circuit-switched networks with *adaptive routing*. Accordingly, a key contribution of this paper is to incorporate an ML technique in the LL routing algorithm and to outperform the conventional LL approach. To this end, we extend the supervised naïve Bayes classifier based on network snapshots that record the historic network state information. This is then used to decide the future connection blocking probability of the entire network if a service is set up on a candidate route between a pair of nodes. Then, this is added to the strategy used to decide the best candidate route to serve a service connection request. Our results show that this ML-assisted LL algorithm significantly outperforms conventional LL routing. To enable the learning process based on a large number of network snapshots, we also develop a parallel computing framework to implement parallel learning and performance evaluation. In addition, a network control system supporting naïve Bayes classifier-assisted LL routing algorithm is also described.

The remainder of this paper is organized as follows. In Section II, we introduce the machine learning-assisted LL routing algorithm, where we include the motivation for introducing ML, introduce the supervised naïve Bayes classifier, and describe how this can be used to enhance the LL routing algorithm. In Section III, we describe a parallel computing system and principle for fast learning under a large number of network snapshots. Section IV presents the network control system for implementing the proposed ML-based LL routing algorithm. We evaluate the performance of the proposed approach in Section V, where the performance of the different routing schemes is compared and analyzed. We conclude the paper in Section VI.

## II. MACHINE LEARNING-ASSISTED LEAST LOAD ROUTING ALGORITHM

Machine learning techniques can be broadly divided into three main categories, namely, supervised learning, unsupervised learning, and reinforcement learning. In this study, we employ the supervised naïve Bayes classifier [34], which belongs to the category of supervised learning, to enhance the performance of the LL routing algorithm.

We first introduce the motivation for developing our ML-assisted LL routing algorithm. We then describe the basic concept of the supervised naïve Bayes classifier and apply this ML technique to assist LL routing for better performance.

*A. Motivation of Enhancing LL Routing Algorithm*

Although the LL algorithm is the most efficient to date for routing in circuit-switched networks, it can still be improved and in certain circumstances, its inefficiency may lead to poor performance. We illustrate this by the network example illustrated in Fig. 1 to show that there is still scope for improving it further.

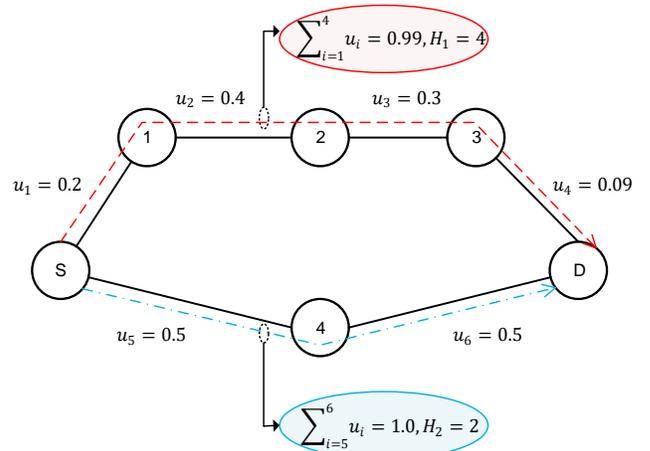

Fig. 1. An example of least loaded routing.

Consider the network of Fig. 1, where the capacity utilization of link $i$ is denoted by $u_i$. For a connection request between nodes $s$ and $d$, the LL routing algorithm would consider the link capacity utilization as the cost of each link to search for a path between the node pair with the smallest sum cost. Accordingly, in this example, the LL routing selects the route (S-1-2-3-D) with a sum cost of 0.99. However, there is a shorter route (S-4-D) with a trivially higher sum cost of 1.0. In this case, the selected LL route is longer than the second route and would consequently consume more link capacity overall to establish the service connection. This presents a dilemma on whether we should choose the LL route with the least congestion or the shorter route with less overall capacity usage. We address this dilemma by using ML-assisted LL routing to improve performance for the connection blocking probability over what would otherwise be achievable by the simple LL routing algorithm with the smallest sum cost.

*B. Supervised Naïve Bayes Classifier*

In machine learning, naïve Bayes classifiers are a family of simple probabilistic classifiers based on Bayes' theorem with strong (naive) independence assumptions between the features. It is capable of calculating the probabilities of different output classes for each input instance.

Let a vector $\boldsymbol{X}$ represent a problem instance, and $x_1, \dots, x_n$ represent *n* features of an instance. Assume that there are *k* class labels, $C_1, \dots, C_k$. Naïve Bayes is a conditional probability model, represented by $P(C_k|\boldsymbol{X})$, which denotes the instance probability of $C_k$ given $\boldsymbol{X}$. Using Bayes' theorem, we can express this conditional probability as

$$P(C_k|\boldsymbol{X}) = P(C_k|x_1, \dots, x_n) = \frac{P(C_k) \cdot P(x_1, \dots, x_n|C_k)}{P(x_1, \dots, x_n)} \quad (1)$$

The "naive" conditional independence assumption assumes that each feature is conditionally independent from other features, and the instance probability can therefore be written as

$$P(C_k|\boldsymbol{X}) = \frac{P(C_k) \cdot \prod_{j=1}^{n} P(x_j|C_k)}{\prod_{j=1}^{n} P(x_j)} \quad (2)$$

*C. Supervised Naïve Bayes Classifier-Based Least Load Routing Algorithm*

We propose to improve the routing performance of the LL routing algorithm by using the supervised naïve Bayes classifier. This is used to predict the future connection blocking probability of the entire network if a service connection is established on a candidate route between a pair of nodes, according to the current capacity utilization status of the network. Based both on this predicted information and the traffic load on each candidate route, we select the route that has the best combination of both low load and small impact on future service connection establishment. We expect that this will help the LL routing algorithm to further reduce the overall service connection blocking probability in the network.

*1) Network snapshot*

In the supervised naïve Bayes classifier, a vector $\boldsymbol{X}$ represents a problem instance. Here we define the network snapshot as a part of the problem instance of the classifier. This network snapshot records the information on the capacity status of each network link. Whenever there is a new service request between a pair of nodes, we record the current network link capacity status as a *snapshot*. With time, we can form a sequence of network snapshots as shown in Fig. 2. In each snapshot, the information on the total link capacity and capacity used on each link is recorded. For example, in Fig. 2, at $t_i$ there is a service request arriving and the current link state is recorded as ($W_j^{(t_i)}, U_j^{(t_i)}$), where $W_j^{(t_i)}$ denotes the total capacity on link *j* and $U_j^{(t_i)}$ denotes the capacity used on link *j*.

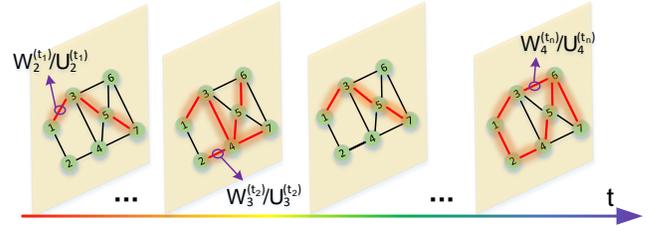

Fig. 2. Network snapshots with time.

Though it is possible for the link total capacity $W_j^{(t_i)}$ to change with time, we assume for simplicity that it is fixed. In particular, we assume that the network snapshot is represented by the vector

$$S^{(i)} = \left[U_1^{(i)}, \cdots, U_j^{(i)}, \cdots, U_L^{(i)}\right]^T \quad (3)$$

Here the superscript *i* denotes the $i^{th}$ network snapshot, which corresponds to the $i^{th}$ service connection request, *L* is the total number of network links, and $U_j^{(i)}$ is the total number of capacity units (e.g., circuits, time slots, or wavelengths) used on link *j*. $U_j^{(i)}$ can be considered as a feature $x_j$ in vector $\boldsymbol{X}$. The assumption of fixed capacity values is justifiable in practice at least for a sufficiently long time-period relevant for the routing design

*2) Predicting blocking probability for a service request between a pair of nodes*

Given a network snapshot (which includes the information on the network link capacity usage) when a new service request between pair of nodes arrives, we use the supervised naïve Bayes classifier to predict its blocking probability. For this, we first define the problem instance (or vector) to be classified as

$$\boldsymbol{X} = [\boldsymbol{S}, sd]^T \quad (4)$$

Here $\boldsymbol{S}$ denotes a network snapshot defined as in (3) and $sd$ is the index of the node pair requesting a service connection; note that the latter is also to be included as a feature of the vector $\boldsymbol{X}$. This new request may be either established or declined depending on the current network capacity utilization. Let the binary variable $Y$ denote the outcome of the classifier when it is given the input $\boldsymbol{X}$. Specifically, $Y = 0$ if the call request between $sd$ is served (i.e., a route is found and allocated to the request); otherwise, $Y = 1$ if the call request between $sd$ is blocked.

We define the conditional probability $P(Y = 1|\boldsymbol{X}) = P(Y = 1|\boldsymbol{S}, sd)$ as the potential blocking probability of a service request between node pair $sd$ based on the current network snapshot $\boldsymbol{S}$. Using Bayes' theorem, this conditional probability can be written as

$$P(Y = 1|\boldsymbol{X}) = \frac{P(Y=1) \cdot P(\boldsymbol{X}|Y = 1)}{P(\boldsymbol{X})} \quad (5)$$

For a circuit-switched network where connection requests arrive dynamically, $P(Y = 1)$ denotes the overall network-wide blocking probability of the service requests that have arrived. Let *H* be the total number of requests that have arrived and $I\{Y^{(i)} = 1\}$ is an indicator function, which equals one if the $i^{th}$ arrived request is blocked, then

$$P(Y = 1) = \frac{\sum_{i=1}^{H} I\{Y^{(i)}=1\}}{H} \quad (6)$$

This effectively calculates the ratio of blocked service requests to the total number of requests that have arrived.

We now consider the probability $P(X|Y = 1)$ which is the probability of finding the network in state $S$ with a request initiated between node pair $sd$ given that the service request is blocked. Using suitable independence assumptions, this may be expanded as follows.

$$P(X|Y = 1) = P(S|Y = 1) \cdot P(sd|Y = 1) = \prod_{j=1}^{L} P(U_j = U_j^S|Y = 1) \cdot P(sd|Y = 1) \quad (7)$$

Here we assume that the capacity usage on the links are independent of each other, and that they are also independent of the node pair that initiates the request. $U_j^S$ is the number of capacity units used on link $j$ in the network snapshot $S$, and $P(U_j = U_j^S|Y = 1)$ is the probability that the number of capacity units used on link $j$ is equal to $U_j^S$ when a service request is blocked.

We can now calculate $P(U_j = U_j^S|Y = 1)$ and $P(sd|Y = 1)$ using the following two equations.

$$P(U_j = U_j^S|Y = 1) = \frac{\sum_{i=1}^{H} I\{U_j^{(i)} = U_j^S \wedge Y^{(i)} = 1\}}{\sum_{i=1}^{H} I\{Y^{(i)} = 1\}} \quad (8)$$

$$P(sd|Y = 1) = \frac{\sum_{i=1}^{H} I\{sd^{(i)} = sd \wedge Y^{(i)} = 1\}}{\sum_{i=1}^{H} I\{Y^{(i)} = 1\}} \quad (9)$$

In (8), $I\{U_j^{(i)} = U_j^S \wedge Y^{(i)} = 1\}$ is an indicator function which is equal to 1 if the $i^{th}$ service request is blocked and the number of capacity units used in link $j$ is equal to $U_j^S$. In (9), $I\{sd^{(i)} = sd \wedge Y^{(i)} = 1\}$ is an indicator function which is equal to 1 if the $i^{th}$ service request is initiated by node pair $sd$ and it is blocked. The denominator in both (8) and (9) is the total number of connections out of $H$ which were blocked

We also need to find $P(X)$ for use in (5), which is derived as follows.

$$P(X) = P(S, sd) = P(S) \cdot P(sd) = \prod_{j=1}^{L} P(U_j = U_j^S) \cdot P(sd) \quad (10)$$

where $P(U_j = U_j^S)$ and $P(sd)$ are calculated as follows.

$$P(U_j = U_j^S) = \frac{\sum_{i=1}^{H} I\{U_j^{(i)} = U_j^S\}}{H} \quad (11)$$

$$P(sd) = \frac{\sum_{i=1}^{H} I\{sd^{(i)} = sd\}}{H} \quad (12)$$

Here $I\{U_j^{(i)} = U_j^S\}$ is an indicator function which is equal to 1 if the number of capacity units used on link $j$ is equal to $U_j^S$ upon the arrival of the $i^{th}$ service request, and $I\{sd^{(i)} = sd\}$ is an indicator function which is equal to 1 if the $i^{th}$ request is initiated by node pair $sd$.

*3) Naïve Bayes classifier-assisted LL routing*

Having explained how to use the naïve Bayes classifier to predict the blocking probability of a service request between a pair of nodes, we next describe an enhanced LL routing algorithm assisted with the supervised naïve Bayes classifier.

Assume that there is a new service request arriving between a node pair $sd$ and at this moment, we have a network snapshot $S$ which records the capacity usage on each link. We also assume that between the node pair $sd$, multiple candidate routes are available for establishing this service connection. The issue then would be *which route should be selected for service connection establishment*. The conventional LL routing algorithm would always choose the one with the least load, even though it may lead to capacity over-consumption under some circumstances (and the consequent higher blocking for subsequent connection requests). Here the term *capacity over-consumption* is defined as the amount of extra capacity required by a selected LL path compared to the shortest path otherwise chosen. For example, if the shortest path of a service connection has two hops, while its selected LL path has three hops, then the capacity over-consumption in this scenario is one unit if the service connection uses one unit of bandwidth. In the naïve Bayes classifier-assisted LL routing algorithm, in addition to the traffic load on a route, we also consider how the establishment of a connection on this route affects the successful establishment of future service connections from the perspective of the service blocking probability of the overall network. The naïve Bayes classifier-assisted LL routing algorithm then aims to achieve the best balance between the two objectives.

Specifically, for each service request between node pair $sd$, we first find the complete set of candidate routes that are eligible for establishing the connection, $R^{sd}$. For this, we first calculate offline, all possible routes between all pairs of nodes based on the network topology. Using this, for each online service request received, we remove all the routes that do not have sufficient remaining capacity from the set of all possible paths between node pair $sd$. Next, we find how the establishment of future service connections would be affected if a service connection is indeed established on a particular route in $R^{sd}$. Here for each candidate route $r_k^{sd} \in R^{sd}$, we first assume that we use it to establish a service connection, after which the network snapshot $S_c$ would be updated to

$$S_k = S_c + r_k^{sd} \quad (13)$$

Then based on $S_k$, we estimate the potential service blocking probability between any node pair $s'd'$ after a service connection is established along $r_k^{sd}$ as

$$BP_{s'd'}^{sd,k} = P(Y = 1|S_k, s'd') \quad (14)$$

This blocking probability can be calculated by equations (5)-(12).

We can then calculate a network-wide blocking probability after the service connection is established on $r_k^{sd}$ as

$$BP_{net}^{sd,k} = \sum_{s'd'} l_{s'd'} \cdot BP_{s'd'}^{sd,k} \quad (15)$$

Here $l_{s'd'}$ is the ratio of traffic load between node pair $s'd'$ to the total traffic in the entire network. The relationship $\sum_{s'd'} l_{s'd'} = 1$ holds and $l_{s'd'}$ is calculated as

$$l_{s'd'} = \frac{\sum_{i=1}^{H} I\{s'd'^{(i)} = s'd'\}}{H} \quad (16)$$

where $I\{s'd'^{(i)} = s'd'\}$ is an indicator function to tell if the $i^{th}$ service request is initiated by node pair $s'd'$. Obviously, the route that has the smallest $BP_{net}^{sd,k}$ should have more preference to establish a service connection since its establishment would result in the lowest blocking probability for the future service connections.

In addition to $BP_{net}^{sd,k}$, for each route $r_k^{sd} \in R^{sd}$, we also estimate their sum load on their traversed links. Specifically, this sum load is

$$u_k^{sd} = \sum_{i \in r_k^{sd}} u_{k,i}^{sd} \quad (17)$$

where $u_{k,i}^{sd}$ is the capacity utilization on the $i^{th}$ link of route $r_k^{sd}$, which is defined as

$$u_{k,i}^{sd} = \frac{U_i^c}{W_i} \quad (18)$$

where $W_i$ is the number of total capacity units on link $i$ and $U_i^c$ is the number of capacity units used on link $i$ in the network snapshot $S_c$. A route with the smallest $u_k^{sd}$ is considered the least congested or having the least load and we should establish a service connection along this route to avoid network congestion.

The naïve Bayes classifier-assisted LL routing considers both the load on each route as well as how the establishment of a service connection on a route would affect the successful establishment of future service connections in the network. Specifically, we select the route based on the following equation.

$$k_{sd}^* = \arg\min_k \left( BP_{net}^{sd,k} \cdot u_k^{sd} \right) \quad (19)$$

which chooses a route with the least load as well as having the lowest impact on the success of future service connection establishment.

*4) Online learning algorithm*

To support online learning, the naïve Bayes classifier-assisted LL routing algorithm can be implemented as follows.

```
Repeat
{
   For a new service request between node pair sd, Z^sd
   {
      Decide its eligible candidate route set R^sd based on all the routes found offline for the node pair;
      For each route r_k^sd ∈ R^sd
      {
         Calculate its sum load u_k^sd using (17), (18);
         Calculate the potential network-wide blocking probability BP_net^{sd,k} after Z^sd is established along r_k^sd, using (13)-(16); specifically, in (14), P(Y = 1|S_k, s'd') is calculated using (5)-(12);
      }
      Choose route k_sd^* = arg min_k (BP_net^{sd,k} · u_k^sd);
      If k_sd^* = NULL, the service request is blocked; otherwise, establish the service request along the route;
      Update all the related parameters in (5)-(12) accordingly.
   }
}
```

In the above learning process, Laplacian smoothing is required when initially calculating some terms. This is performed as follows.

$$P(Y = 1) = \frac{\sum_{i=1}^{H} I\{Y^{(i)}=1\}+1}{H+2} \quad (20)$$

$$P(U_j = U_j^S | Y = 1) = \frac{\sum_{i=1}^{H} I\{U_j^{(i)}=U_j^S \wedge Y^{(i)}=1\}+1}{\sum_{i=1}^{H} I\{Y^{(i)}=1\}+W_j+1} \quad (21)$$

$$P(sd|Y = 1) = \frac{\sum_{i=1}^{H} I\{sd^{(i)}=sd \wedge Y^{(i)}=1\}+1}{\sum_{i=1}^{H} I\{Y^{(i)}=1\}+m} \quad (22)$$

$$P(U_j = U_j^S) = \frac{\sum_{i=1}^{H} I\{U_j^{(i)}=U_j^S\}+1}{H+W_j+1} \quad (23)$$

$$P(sd) = \frac{\sum_{i=1}^{H} I\{sd^{(i)}=sd\}+1}{H+m} \quad (24)$$

$$u_{k,i}^{sd} = \frac{U_i^c}{W_i} + \alpha \quad (25)$$

where $W_j$ is the total number of capacity units on link $j$, $m$ is the total number of node pairs in the network, and $\alpha$ a small value set as $10^{-6}$.

Also, in the course of service provisioning and learning, all the system parameters in (5)-(12) are updated upon each arrived service request, no matter eventually served or blocked.

## III. PARALLEL DYNAMIC-EVENT SIMULATIONS FOR FAST LEARNING

We have designed a parallel learning system for the machine learning process. Fig. 3 shows our parallel learning system, which consists of a cluster controller and computing resources. The cluster controller is the main component of the parallel computing system, which is responsible for distributing the computation tasks to the computing resources and gathering the results fed back by the computing resources. The computing resources are a cluster of computers, which work in parallel and feedback the results of their computations to the cluster controller. We have constructed a prototype of this parallel learning system in our laboratory as shown at the two bottom corners of Fig. 3. The system contains 10 mini-computers, one Ethernet switch, one power supply module, and three fans. Each mini-computer has a 4.0-GHz quad-core CPU and 8-GB memory [35].

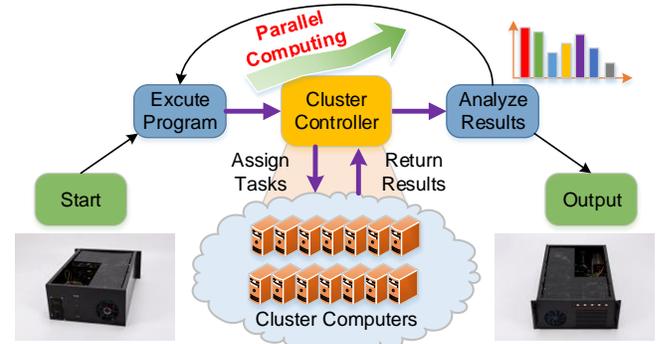

Fig. 3. Parallel learning system.

Fig. 4 shows how the parallel computing system implements the parallel simulation and learning of the naïve Bayes classifier-assisted LL routing algorithm. For each dynamic-event simulation and learning process, the cluster controller splits the simulation into a series of sub-simulations according to the number of call arrival requests. For example, consider running a simulation and learning process with 100 million service arrival events. We first generate 10 sub-simulation tasks with each simulating one million arrival events. The cluster controller will distribute each one-million-event sub-simulation to a particular cluster computer for parallel computing. We assume that there are ten such cluster computers. Upon receiving the sub-simulation tasks, each cluster computer will start the simulation and learning process independently and in parallel. In addition to counting the number of blocked events, they also learn network snapshots and update the related parameters in (5)-(18). Once a cluster computer completes the one-million-event sub-simulation, it feedbacks the number of blocked events as well as those learned parameters to the cluster controller.

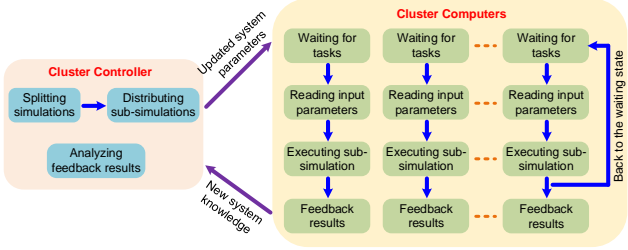

Fig. 4. Procedure of parallel computing.

Once the cluster controller receives this information, it would update the total number of blocked events in the entire simulation system and the related parameters in (5)-(18) by integrating all the feedback information from all the cluster computers. For example, $\sum_{i=1}^{H} I\{U_j^{(i)} = U_j^S \wedge Y^{(i)} = 1\}$ counts the total number of situations when the $i^{th}$ service request is blocked and the number of capacity units used on link $j$ is just equal to $U_j^S$ in an H-arrival-event simulation. For each cluster computer, after simulating one million arrival events, we can find its corresponding number $s_n^k = \sum_{i=1}^{H=10^6} I\{U_j^{(i)} = U_j^S \wedge Y^{(i)} = 1\}$, where $n$ is the index of the cluster computer in the parallel system and $k$ is the round index of the sub-simulations that the cluster computer executes. Thus, for each round of sub-simulation tasks, with the cluster computers executing the tasks in parallel, we will have a corresponding sum parameter at the $k^{th}$ round as $s^k = \sum_{n \in Cluster} s_n^k$, where $Cluster$ denotes the set of cluster computers. In a similar way, we can update other related parameters for the learning process.

In order to learn the whole routing system as much as possible, we need to have multiple rounds of the sub-simulations and learning processes. Continuing with the previous example, as the total number of simulated arrival events is 100 million and each round of sub-simulations performed by the parallel system can simulate 10 million arrival events, we need to run 10 rounds of sub-simulations. Each time when a round of sub-simulations is completed, the cluster controller will initiate the second round of sub-simulations. When doing this, the cluster controller would also forward the parameters in (5)-(18) learned in the previous rounds to each of the cluster computers. The latter will use these updated parameters to run simulations for another round of arrival events and meanwhile collect the information for updating the related parameters. The entire process will terminate when the total number of planned simulation events are reached. Then the cluster controller will find the final service connection blocking probability and the final updated parameters in (5)-(18).

## IV. IMPLEMENTATION OF ML-ASSISTED NETWORK CONTROL SYSTEM

In Section II, we described the mathematical fundamentals of applying the naïve Bayes classifier to assist the LL algorithm. In Section III, we showed how to generate various artificial arrival events and network snapshots for the learning system to learn based on forecast traffic load between different node pairs, in order to have sufficient network snapshots for learning. In this section, we introduce how this ML-assisted routing algorithm may be applied in a real operational network, such as the one shown in Fig. 5, to handle real time demand requests.

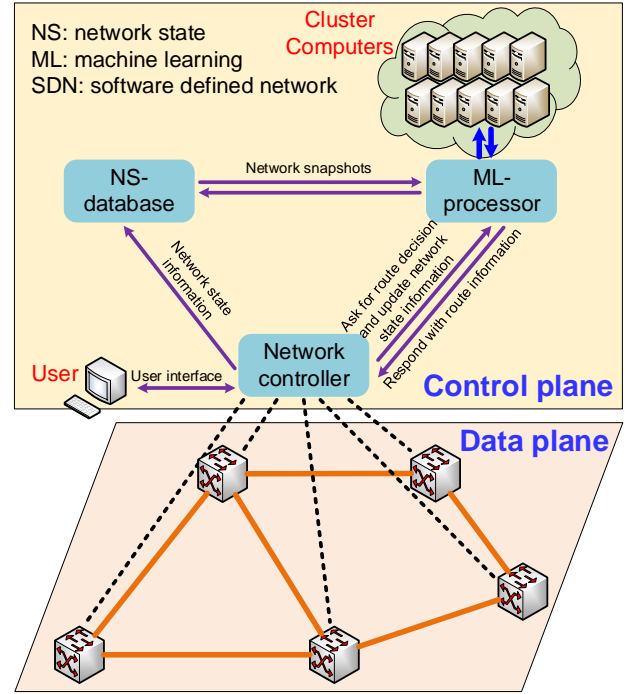

Fig. 5. Diagram of a real network control system.

For this, we assume a centralized network control system with a single network controller. This assumption is reasonable since today a Software Defined Network (SDN) control system is implemented in just such a centralized fashion. In addition to providing an interface to its users, for service applications, the central controller monitors and records the state information of the entire network; it also receives service requests from users for the establishment of service connections according to the current network status. When provisioning a service connection, it will consult its ML-processor to provide the information on whether the request should be blocked or decide the best route used to establish a service connection if the service request is accepted.

The ML-processor is responsible for two tasks. The first task is to implement initial learning process based on a simulation model of the actual network and the forecasted traffic load matrix, as described in Section II. After this initial learning, the ML-processor can provide an initial set of learning parameters using (5)-(18). Based on this, it can then undertake the second task to make a decision on whether a service connection can be provisioned and decide its route while continuing to carry out online learning based on real network service request data and network snapshot information, i.e., by updating the related learning parameters in (5)-(18).

The ML-processor will feedback the decision information to the network central controller and the latter will trigger the required signaling to establish service connections. Meanwhile, it will also send the information on the current network snapshot and service connection to a network state database (NS-database) for storing the network's historic state information. Note that with the accumulation of this information, there would be more real network state data for the ML-processor to learn.

For the ML-process, in addition to the initial offline learning, other offline learning may also happen under some situations. For example, if we can predict that there would be a clear change of traffic load between a pair of nodes in the near future, then we can add that to the learning process to tune the related learning parameters in (5)-(18). This will be done by jointly considering the simulation-generated artificial data and the real network state information data stored in the NS database. Similarly, if there is an upgradation of network link capacity, we need to implement a similar re-learning process to update those related learning parameters. Since the ML-processor needs to carry out significant computing for the learning based on artificial simulations and the real network state information data stored in the NS database, a parallel computing system such as the one shown in Section III would be desirable for this purpose.

In addition, we need to consider the possibility of service release after it completes its mission. When this happens, the central controller will instruct the data plane to release the related network resources and inform its ML-processor to update its current network state information. However, for the service release process, the related learning parameters do not need to be updated and no network snapshots need to be forwarded to the NS-database for storage.

## V. SIMULATIONS AND PERFORMANCE ANALYSES

### A. Test Conditions

We assume that new service connection requests arrive following a Poisson arrival process with an arrival rate of $\lambda$ per second. Each service connection request has a mean holding time of $1/\mu$ seconds following an exponential distribution. We normalize our time measurement using $1/\mu = 1$ so that the traffic load between each pair of nodes may be considered to be $\lambda$ erlang. For all the study schemes, they used the same random seed to generate the randomly arrived service requests following the same random process. We consider two test networks: (1) the 14-node, 21-link NSFNET network and (2) the 21-node, 25-link ARPA-2 network as shown in Fig. 6. In the NSFNET network, the number of capacity units in each link is random and is uniformly distributed within the range of [5, 27]. The traffic load between each node pair is generated randomly with a uniform distribution within the range of $[0.45, 0.45 + X]$ erlang, where $X \in \{0.15, 0.3, ..., 1.05\}$. In the ARPA-2 network, the number of capacity units in each link is similarly randomly distributed within a range of [5, 31] and the traffic load between each node pair is generated within a range of $[0.07, 0.07 + X]$ erlang, where $X \in \{0.07, 0.14, ..., 0.49\}$.

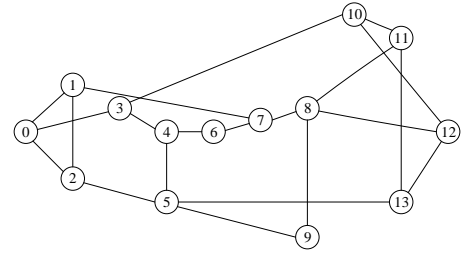

(a) 14-node and 21-link NSFNET network.

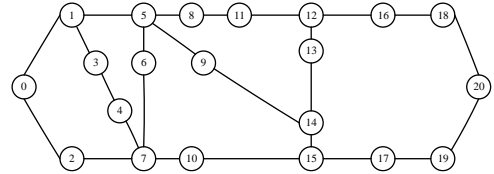

(b) 21-node and 25-link ARPA-2 network.

Fig. 6. Test networks.

For performance comparison, we have also run simulations based on the adaptive shortest path routing algorithm and the conventional LL routing algorithm. The shortest path routing algorithm always finds an eligible route with the smallest number of hops based on the current network capacity usage. This is adaptive shortest path routing, not a fixed one. The conventional LL routing algorithm uses the link capacity utilization as the "cost" of each link, and then finds a feasible route with the smallest $u_k^{sd}$ in (17) by using Dijkstra's algorithm. Both of the algorithms try to find their own best eligible routes between a pair of nodes. If no such routes can be found, the service request would be blocked and discarded. Each blocking probability point of the shortest path routing and the conventional LL algorithms is evaluated based on $10^6$ service connection request arrival events.

For the naïve Bayes classifier-assisted LL routing algorithm, we employ the parallel computing system (as described in Section III) to run dynamic-event driven simulations with $10^8$ call arrival requests. Between each node pair, all the eligible routes are considered for the route selection in the algorithm based on the current network capacity utilization status. We employed the parallel computing system containing ten min-computers as shown in Fig. 3 to run the simulation and learning process, in which one computer functions as a cluster controller as well as a cluster computer and all the other computers function as cluster computers. Each computer has a 4.0-GHz quad-core CPU and 8-GB memory. Ten rounds of sub-simulations are executed with each round simulating 10 million service arrival events and one computer simulating one million service arrival events. The overall blocking probability of the naïve Bayes classifier-assisted LL routing algorithm was calculated based on the total $10^8$ call arrival requests.

### B. Service Connection Blocking Performance

We first compare the service connection blocking probability for the different routing algorithms. Fig. 7 shows the blocking probabilities of the NSFNET and ARPA-2 networks with an increasing interval of random traffic load per node pair *X*. In the legend, "LLA" corresponds to the conventional LL routing algorithm, "SP" corresponds to the adaptive shortest path routing algorithm, and "ML-NB-LL" corresponds to the naïve Bayes classifier-assisted LL routing algorithm.

Fig. 7(a) shows the results of the SP and conventional LL algorithm for the NSFNET network with 95% confidence intervals based on Student-t distribution. For the naïve Bayes classifier-assisted LL routing algorithm, we do not show its confidence interval due to its extremely large number of learned network snapshots (i.e., up to $10^8$ arriving requests). We can see that the naïve Bayes classifier-assisted LL routing algorithm achieves the least blocking probability among all the three routing algorithms, the conventional LL routing algorithm ranks second in terms of blocking probability, and the shortest path routing algorithm performs worst. It is reasonable that the conventional LL algorithm outperforms the SP algorithm since the former aims for balancing the load in the whole network by always choosing a route with the least congestion. Meanwhile, the naïve Bayes classifier-assisted LL routing algorithm can outperform the conventional LL algorithm because in addition to the traffic load on each route, it also considers the future potential network-wide blocking probability predicted based on the historic service provisioning data when a service connection is established on a specific route. By considering this data, the LL algorithm chooses a lightly loaded route that also has the least adverse impact on future service connection establishment if a service connection is established on this route. A similar observation can be made for the ARPA-2 network (see Fig. 7(b)) where the performance ranking of the three algorithms in terms of blocking probability is the same.

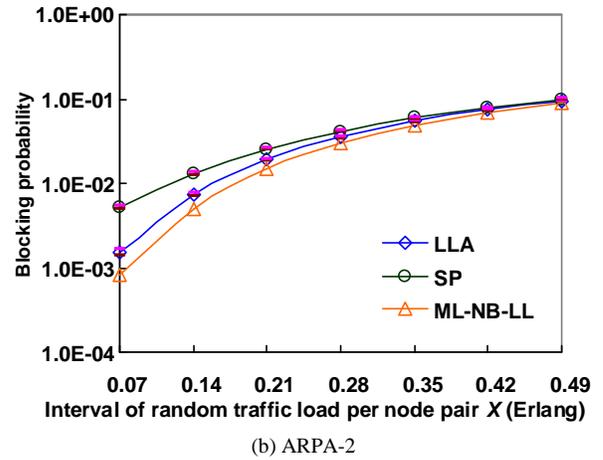

(b) ARPA-2

Fig. 7. Blocking performance comparison between the three routing algorithms.

Today, researchers have generally reached a consensus that the LL or least congested routing algorithm is the method of choice and therefore it can be used as a benchmark for routing methods aiming at minimizing blocking probability. However, as shown by our example (see Fig. 1), we observe certain inefficiency in the LL routing algorithm that might potentially suffer from resource over-consumption in certain situations. In this study, with the assistance of a machine learning technique, i.e., the naïve Bayes classifier, we demonstrate that the blocking performance of the LL algorithm can be improved by learning from the historic network service provisioning data (i.e., network snapshots).

### C. Impact of Number of Snapshots Learned

Learning network snapshots is an important step to ensure a good performance for the naïve Bayes classifier-assisted LL routing algorithm. Thus, we also evaluated how the number of network snapshots learned (which is equal to the number of arrived service requests) can impact the service connection blocking performance. The results are shown in Fig. 8, where the legend "Single" corresponds to the case of a single computer for the simulation, the legend "Para" corresponds to the case of parallel computing, and the legend "Time" corresponds to the time required for simulating a certain number of arrival events.

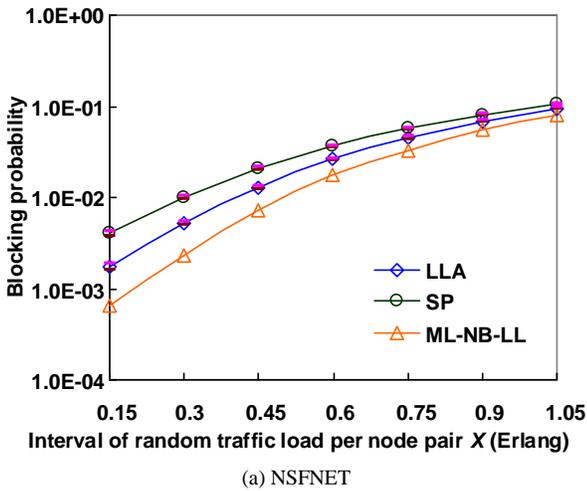

(a) NSFNET

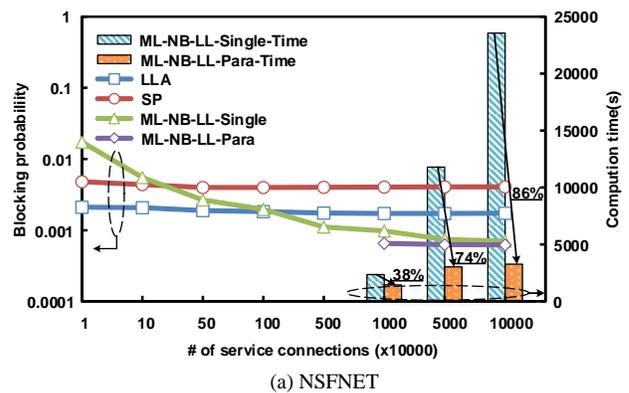

(a) NSFNET

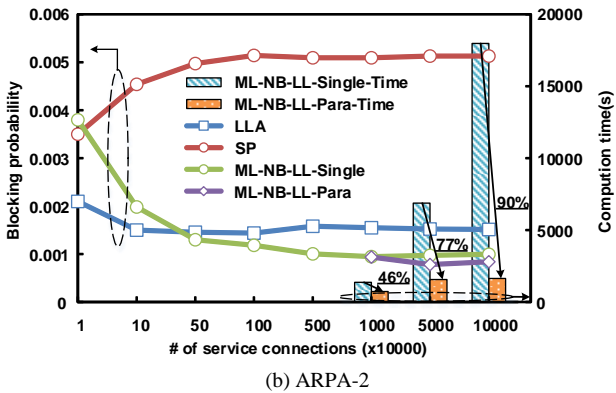

(b) ARPA-2

Fig. 8. Blocking performance and computing time comparison between the different routing algorithms under different numbers of network snapshots learned.

In the NSFNET network, the traffic load between each node pair is randomly generated within the range of [0.45, 0.6] erlang, and the total number of capacity units on each link is randomly distributed within the range of [5, 27]. For the SP and conventional LL algorithms, as there is no a learning process, their blocking performances almost do not change after one million service arrival events. Nonetheless, for the naïve Bayes classifier-assisted LL routing algorithm, we see that with an increasing number of service connection requests, i.e., the number of learned network snapshots, its blocking performance is improved and then almost does not change when 50 million arrival events are simulated. Moreover, the final blocking performance of the naïve Bayes classifier-assisted LL routing algorithm is better than that of the other two routing algorithms.

In addition, specifically for the naïve Bayes classifier-assisted LL routing algorithm, because there are up to 100 million arrival events to simulate, which is very time consuming for a single computer, parallel computing system has also been employed for the simulation. In Fig. 8(a), we also compare the times consumed by a single computer and the parallel computing system. We see that the simulation time can be significantly shortened by the parallel computing system. For example, for the simulation of 100 million arrival events, the computation time can be reduced by over 85% with the parallel computing system, from almost 7 hours to about 1 hour. This indicates the computational efficiency of the parallel computing system for the machine learning process. Note that, in the future, we may further increase the number of computers in the parallel computing system for an even shorter computation time since our system can be easily expanded.

We have conducted similar simulation studies for the ARPA-2 network, in which the traffic load between each node pair is randomly generated within the range of [0.07, 0.14] erlang, and the total number of capacity units in each link is randomly distributed within the range of [5, 31]. As shown in Fig. 8(b), we have similar observations of the blocking performance and the computation times for the different schemes to those of the NSFNET network. Here, for the naïve Bayes classifier-assisted LL routing algorithm, the blocking performance almost does not change when 10 million arrival events are simulated. Also, the naïve Bayes classifier-assisted LL routing algorithm can always achieve better performance than the other two routing algorithms. We see that the parallel computing technique can help to reduce up to 90% computation time for the machine learning process.

### D. How Does Naïve Bayes Classifier Help to Avoid Network Capacity Over-Consumption for the LL Algorithm?

The improved performance of the naïve Bayes classifier-assisted LL routing algorithm is attributed to its learning capability from the historic network service provisioning data (i.e., network snapshots). As a result, this effectively controls the network capacity over-consumption that the conventional LL algorithm may suffer from in some situations. In the following, we demonstrate by our analysis and simulations, how this has been achieved.

In Fig. 9, we show the *extra hop count* for each established service connection compared to their shortest paths. For example, suppose that the shortest path between the node pair of a service connection (based on the physical topology) has $K$ hops. However, considering the network resource usage status, the final selected route by the LL routing algorithm is not the shortest one, but a longer one. The number of hops $\Delta$ by which this is larger than $K$ is defined as the *extra hop count* for the route compared to the shortest path. The hop count of this selected route is $K + \Delta$. Thus, in Fig. 9, $\Delta = 0$ means the case of the shortest path and $\Delta > 0$ means that the selected route is longer than the shortest path. We compare the hop count distributions of established service connections for the shortest path routing algorithm, the conventional LL routing algorithm, and the naïve Bayes classifier-assisted LL routing algorithm. For the former two, only $10^6$ arriving requests are simulated owing to its stabilization of blocking probability after this number of arriving request as shown in Fig. 9. For the last one, we simulated $10^8$ arrival requests and collected the above distribution data only for the last $10^6$ arrival requests of the entire simulation.

According to the results in Fig. 9, there are dominant percentages of service connections provisioned on their shortest routes with $\Delta = 0$. However, there are also large percentages of service connections provisioned on the second and third shortest routes. In some cases, very long routes with $\Delta = 10$ could be used under the LL routing algorithms since they only consider the least congested routes while ignoring the actual hop counts of the selected routes. Of course, the LL algorithms also by default partially consider the hop count, which is weighted by the link capacity utilization. We also calculated the average number of extra hops for the provisioned service connections for the three routing algorithm (see the sub-captions of Fig. 9). It can be found that in the NSFNET network, the naïve Bayes classifier-assisted LL routing algorithm shows the smallest number of extra hops compared to the other two routing algorithms. This implies that the naïve Bayes classifier-assisted LL routing algorithm has the lowest network capacity over-consumption when implementing the LL routing algorithm, thereby achieving a lower overall blocking probability. The same comparison can be made for the ARPA-2 network, where similar phenomena can be observed. Most of the service connections were served by the first, second, and third shortest routes between node pairs. The shortest path routing algorithm has the smallest number of extra hops, while comparing the two LL routing algorithms, the naïve Bayes classifier-assisted LL routing algorithm shows a smaller number of extra hops. This implies that it wastes less network capacity when choosing the LL routes than the conventional LL routing algorithm. Based on these results, we can conclude that naïve Bayes classifier-assisted LL routing algorithm has controlled well the routes' extra hops to avoid over-consumption of network capacity when choosing the LL routes. Consequently, it significantly reduced the service connection blocking probability as earlier shown in Fig. 7.

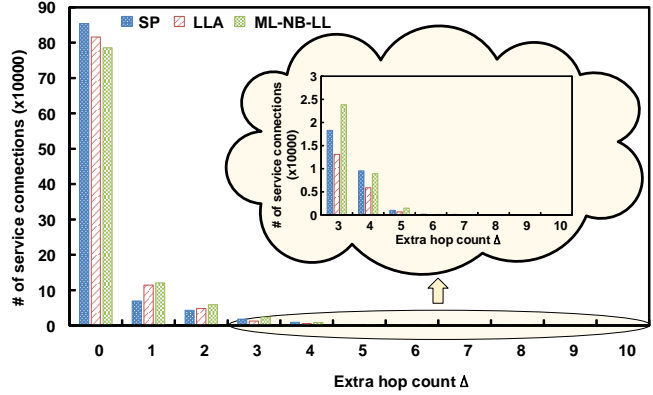

(a) NSFNET (average numbers of extra hops = 0.2576 (SP), 0.2781 (LLA), 0.2194 (ML-NB-LL))

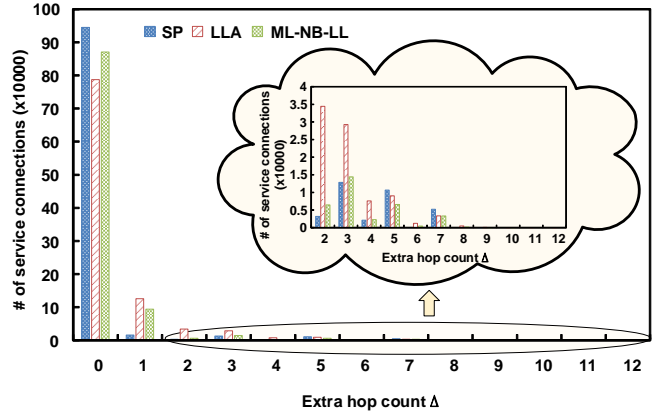

(b) ARPA-2 (average numbers of extra hops = 0.1602 (SP), 0.3942 (LLA), 0.3575 (ML-NB-LL))

Fig. 9. Extra hop count distributions of the different routing algorithms.

### E. Can We Improve Blocking Performance of Conventional LL Routing Algorithm by Simply Uniformly Controlling Extra Hop Counts?

We see that the ML-assisted LL routing algorithm can improve the blocking performance over the conventional LL routing algorithm through *differently* controlling the extra hop count to the shortest path for each established service connection. The question for us is: *can we also improve the blocking performance through simply uniformly setting a limit on the extra hop count to the shortest path for the conventional LL routing algorithm when establishing a lightpath service connection*? We did this through extending the conventional LL routing algorithm by incorporating this extra hop count limit in the algorithm when searching for a least loaded route in a network. We show the related results of blocking performance with the various extra hop counts as shown in Fig. 10. Here we consider the NSFNET network with the capacity units on each link distributed within the range of [5, 27], and with the traffic load per node pair distributed within the range of [0.45, 0.45+$X$] erlang, where $X$ is the interval of traffic load between different node pairs. From the result we can see that with an increasing extra hop count to the shortest path, the lightpath blocking performance is improved at the beginning, which is however saturated after a certain threshold of the extra hop count. More specifically, when the extra hop count grows to 4, a further increase of this parameter would not bring the improvement of blocking performance, till it becomes a full version of conventional LL routing algorithm when the extra hop count becomes infinite. This observation implies that it does not help improve the blocking performance for the conventional LL routing algorithm by simply controlling the extra hop count uniformly for all the node pairs. Instead, such controlling should performed differently for different node pairs according to the current network status as in the naïve Bayes classifier-assisted LL routing algorithm.

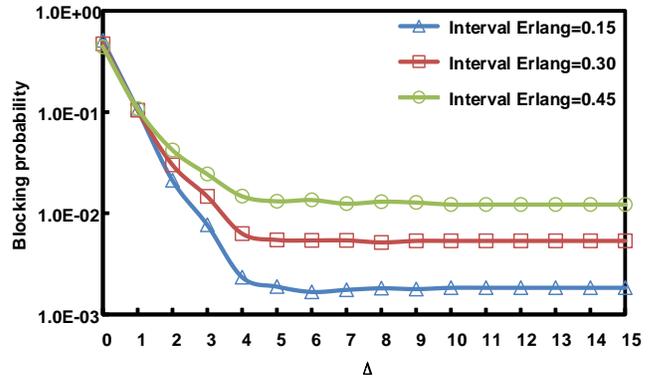

Fig. 10. Blocking probability versus hop count limit for an LL route (NSFNET). Δ: the difference between hop count of an LL route and the count of the shortest route between each node pair.

## VI. CONCLUSION

We have considered the blocking probability benchmark established by the conventional LL routing algorithm. We have demonstrated that this benchmark can be improved because LL routing may over-consume network capacity under certain circumstances when selecting an LL route to provision a service connection. In the past, it has been difficult to find an effective way to overcome this type of inefficiency. In this paper, we employed a machine learning technique to develop a naïve Bayes classifier-assisted LL routing algorithm aiming to overcome the above capacity inefficiency and further break through the blocking performance benchmark of the conventional LL algorithm. We have considered network snapshots as the historic data for machine learning. This has been applied to a supervised naïve Bayes classifier to predict the potential blocking probabilities of future call requests in the whole network if a service request is established along a certain route. With this information, a route that has low load as well as low adverse impact on the success of future service connection establishment. To implement the learning process based on a large number (hundreds of millions of) of network snapshots, we have employed a parallel computing system built in our laboratory to implement parallel learning and performance evaluation. Moreover, the implementation of this ML-assisted network control system was also addressed.

Simulation studies have demonstrated that the proposed naïve Bayes classifier-assisted LL routing algorithm could indeed achieve improvement over the benchmark performance set by the conventional LL routing algorithm. Moreover, we have also investigated the phenomenon of over-consumption of network capacity by the conventional LL routing algorithm. We have demonstrated that the conventional LL routing algorithm has a much larger number of extra hops for service connections provisioning than that of naïve Bayes classifier-assisted LL routing algorithm. Thus, the naïve Bayes classifier-assisted LL routing algorithm is effective in controlling well the extra hops for service connections when provisioned based on the LL routes. Also, for the naïve Bayes classifier-assisted LL routing algorithm, its blocking performance is closely related to the number of network snapshots learned. An increasing number of network snapshots can lead to lower blocking probability. Finally, we have found that the parallel computing system that we employed specifically for learning the required large numbers of network snapshots is computationally efficient. Up to 90% of the time can be saved for the learning process, which significantly accelerated our simulation study.

**Gangxiang Shen** received his B.Eng. degree from Zhejiang University, China; his M.Sc. degree from Nanyang Technological University, Singapore; and his Ph.D. degree from the University of Alberta, Canada, in January 2006. He is a Distinguished Professor with the School of Electronic and Information Engineering of Soochow University in China. Before he joined Soochow University, he was a Lead Engineer with Ciena, Linthicum, Maryland. He was also an Australian ARC Postdoctoral Fellow with University of Melbourne. His research interests include integrated optical and wireless networks, spectrum efficient optical networks, and green optical networks. He has authored and co-authored more than 150 peer-reviewed technical papers. He is a Lead Guest Editor of IEEE JSAC Special Issue on "Next-Generation Spectrum-Efficient and Elastic Optical Transport Networks," and a Guest Editor of IEEE JSAC Special Issue on "Energy-Efficiency in Optical Networks." He is an associated editor of IEEE/OSA JOCN, and an editorial board member of Optical Switching and Networking and Photonic Network Communications. He was a Secretary for the IEEE Fiber-Wireless (FiWi) Integration Sub-Technical Committee. He received the Young Researcher New Star Scientist Award in the "2010 Scopus Young Researcher Award Scheme" in China. He was a recipient of the Izaak Walton Killam Memorial Award from the University of Alberta and the Canadian NSERC Industrial R&D Fellowship.

**Longfei Li** received the master's degree from Soochow University, China, in 2013. He is currently pursuing the Ph.D. degree with the School of Electronic and Information Engineering, Soochow University. His research interests include optical network design, cloud computing, machine learning and optimization.

**Ya Zhang** is currently a Master student with Soochow University in China. His current research interest focuses on optical networks.

**Moshe Zukerman** (M'87–SM'91–F'07) received his B.Sc. in Industrial Engineering and Management and his M.Sc. in Operation Research from Technion-Israel Institute of Technology and a Ph.D. degree in Engineering from The University of California Los Angeles in 1985. During 1986-1997 he served in Telstra Research Laboratories (TRL). During 1997-2008 he was with The University of Melbourne. In Dec 2008, he joined City University of Hong Kong where he is a Chair Professor of Information Engineering. He has served on the editorial boards of various journals such as IEEE JSAC, IEEE/ACM Transactions on Networking, IEEE Communications Magazine, Computer Networks and Computer communications and on numerous conference committees. Prof. Zukerman has over 350 publications in scientific journals and conference proceedings, has been awarded several national and international patents. He has served as a member and Chair of the IEEE Koji Kobayashi Computers and Communications Award Committee.

**Sanjay K. Bose** did his B.Tech. from IIT, Kanpur (1976) and his M.S. and PhD. (1977 and 1980) from the State University of New York in Stony Brook. During 1980-1982, he worked in the Corporate Research and Development Center of the General Electric Co. in Schenectady, N.Y. on projects associated with power line communications, optical fiber communications and mobile-satellite communications. He subsequently joined the faculty in the Department of Electrical Engineering, IIT Kanpur in 1982 and worked there until 2003. From 2003 to 2008, he was on the faculty of the School of EEE in NTU, Singapore. He returned to India in 2009 and joined the faculty of the Department of EEE, IIT Guwahati. During 2011-2014, he was also the Dean, Alumni Affairs and External Relations in IIT Guwahati. Prof. Bose has also held short-term and long-term visiting appointments in the University of Adelaide, Queensland University of Technology, Nanyang Technological University and University of Pretoria. More details on Dr. Bose may be found in his home page and related links at the URL http://www.iitg.ernet.in/skbose/